\documentclass[preprintnumbers,nofootinbib,preprint,12pt,showpacs]{revtex4}

\usepackage{amssymb,epsf}

\usepackage{latexsym}

\begin{document}

{~}

\vspace{1cm}

\title{Charged Rotating Kaluza-Klein Black Holes in Dilaton Gravity
\vspace{1cm}
}

\author{
${}^{1}$Masoud Allahverdizadeh,
${}^{2}$Ken Matsuno\footnote{matsuno@sci.osaka-cu.ac.jp}
and
${}^{3}$Ahmad Sheykhi\footnote{sheykhi@mail.uk.ac.ir}
\vspace{0.5cm}
}

\affiliation{
${}^{1}$ Institut f\"{u}r Physik, Universit\"{a}t Oldenburg, D-26111 Oldenburg, Germany
\\
${}^{2}$ Department of Mathematics and Physics, Graduate School of Science,
Osaka City University, 3-3-138 Sugimoto, Sumiyoshi-ku, Osaka 558-8585, Japan
\\
${}^{3}$ Department of Physics, Shahid Bahonar University, Kerman
76175, Iran}

\vspace{1cm}

\begin{abstract}

We obtain a class of slowly rotating charged Kaluza-Klein black
hole solutions of the five-dimensional Einstein-Maxwell-dilaton
theory with arbitrary dilaton coupling constant. At infinity, the
spacetime is effectively four-dimensional. In the absence of the
squashing function, our solution reduces to the five-dimensional
asymptotically flat slowly rotating charged dilaton black hole
solution with two equal angular momenta. We calculate the mass,
the angular momentum and the gyromagnetic ratio of these rotating
Kaluza-Klein dilaton black holes. It is shown that the dilaton
field and the non-trivial asymptotic structure of the solutions
modify the gyromagnetic ratio of the black holes. We also find
that the gyromagnetic ratio crucially depends on the dilaton
coupling constant, $\alpha$, and decreases with increasing
$\alpha$ for any size of the compact extra dimension.

\end{abstract}


\date{\today}

\maketitle 

\newpage

\section{Introduction}
The study of black holes in more than four spacetime
dimensions is motivated by several reasons. Strong motivation comes
from developments in string/M-theory, which is believed to be the most consistent approach
to quantum theory of gravity in higher dimensions. In
fact, the first successful statistical counting of black hole
entropy in string theory was performed for a five-dimensional
black hole \cite{Stro}. This example provides the best laboratory
for the microscopic string theory of black holes. Besides, the
production of higher-dimensional black holes in future colliders
becomes a conceivable possibility in scenarios involving large
extra dimensions and TeV-scale gravity.
Furthermore, as
mathematical objects, black hole spacetimes are among the most
important Lorentzian Ricci-flat manifolds in any dimension. While
the non-rotating black hole solution to the higher-dimensional
Einstein-Maxwell gravity was found several decades ago \cite{Tan},
the counterpart of the Kerr-Newman solution in higher dimensions,
that is, the charged generalization of the Myers-Perry solution
\cite{Myer} in higher dimensional Einstein-Maxwell theory, still
remains to be found analytically. Indeed, the case of charged
rotating black holes in higher dimensions has been discussed in
the framework of supergravity theories and string theory
\cite{Cvetic0,Cvetic1,Cvetic2}.
Recently, charged rotating black
hole solutions in higher dimensions with a single rotation
parameter in the limit of slow rotation have been constructed in
\cite{Aliev2} (see also \cite{
Aliev3,kunz1,AlievC}).  

On the other hand, a scalar field called dilaton appears in the
low energy limit of string theory. The presence of the dilaton
field has important consequences on the causal structure and the
thermodynamic properties of black holes. Thus, much interest has
been focused on the study of the dilaton black holes in recent
years. While exact dilaton black hole solutions of
Einstein-Maxwell-dilaton (EMd) gravity have been constructed by
many authors (see e.g.
\cite{CDB1,HS,CDB2,Cai,Sheykhi1,Astefanesei:2006sy,Charmousis:2009xr}),
exact rotating dilaton black hole solutions have been obtained
only for some limited values of the dilaton coupling constant
\cite{kun,kunz2,Bri}. For the general dilaton coupling constant,
the properties of  charged rotating dilaton black holes only with
infinitesimally small charge \cite{Cas} or small angular momentum
in four \cite{Hor1,Shi,Sheykhi2} and five dimensions have been
investigated \cite{Sheykhi3}. Recently, charged slowly rotating
dilaton black hole solutions in the background of AdS spaces have
also been constructed in arbitrary dimensions
\cite{Sheykhi4,ShAll,SAcosmo,Setal}.

Most authors have considered mainly asymptotically flat
and stationary higher dimensional black hole solutions since
they would be idealized models if such black holes
are small enough for us to neglect the tension of a brane
or
effects of compactness of extra dimensions.
However, if not so,
we should consider the higher dimensional spacetimes which have
another asymptotic structure.
Therefore, it is also important to study black hole solutions
with a wide class of
asymptotic structures.
Recently, the black object solutions with non-trivial
asymptotic structures
have been studied by various authors.
For example, squashed Kaluza-Klein black hole solutions
\cite{DM,GW,Rash2,GGHPR,GSYhole,IM,IKMTmulti,TW,SSY,BR,NIMT,TIMN,MINT,TI,SSW,TYM,GS,YIKT,Stelea:2009ur}
asymptote to the locally flat spacetime, i.e.,
a twisted $\rm S^1$ fiber bundle over the four-dimensional Minkowski spacetime.
In other words, the spacetime is effectively four-dimensional at
the infinity. The black ring solutions with the same asymptotic
structures were also found
\cite{BKW,EEMRring,GSYring,CEFGS,BDGRW}. As far as we know,
charged Kaluza-Klein dilaton black holes with asymptotically
locally flat structures have been constructed in the static case
only \cite{SSY}.
In the present work we would like to generalize such static dilaton black holes to the rotating ones.   
We also investigate the properties of these rotating black holes related to
the presence of the dilaton field and the difference of the
asymptotic structures. Especially, we want to construct a new class of charged rotating squashed Kaluza-Klein black hole solutions 
in five-dimensional 
EMd gravity.
We shall also study the properties of the solutions in the various
limits. Finally, we investigate the effects of the dilaton field
and the twisted compact extra dimension on
the angular momentum and the gyromagnetic ratio of these rotating 
black holes.

\section{ Rotating dilaton black holes with squashed horizons}
We consider five-dimensional EMd theory with action
\begin{eqnarray}
S &=&\frac{1}{16\pi }\int_{\mathcal{M}} d^{5}x\sqrt{-g}\left(
R\text{
}-2\partial_{\mu}\Phi \partial^{\mu}\Phi-e^{-2 \alpha \Phi}F^{\mu \nu }F_{\mu \nu }\right)   \nonumber \\
&&-\frac{1}{8\pi }\int_{\partial \mathcal{M}}d^{4}x\sqrt{-h
}\Theta (h ),  \label{act1}
\end{eqnarray}
where ${R}$ is the scalar curvature, $\Phi$ is the dilaton field,
$F_{\mu \nu }=\partial _{\mu }A_{\nu }-\partial _{\nu }A_{\mu }$
is the electromagnetic field tensor, and $A_{\mu }$ is the
electromagnetic potential. $\alpha $ is an arbitrary constant
governing the strength of the coupling between the dilaton and the
Maxwell field. The last term in Eq. (\ref{act1}) is the
Gibbons-Hawking boundary term which is
chosen such that the variational principle is well-defined. The manifold $%
\mathcal{M}$ has metric $g_{\mu \nu }$ and covariant derivative
$\nabla _{\mu }$. $\Theta $ is the trace of the extrinsic
curvature $\Theta ^{ab}$ of any boundary $\partial \mathcal{M}$ of
the manifold $\mathcal{M}$, with induced metric $h _{ab}$. The
equations of motion can be obtained by varying the action
(\ref{act1}) with respect to the gravitational field $g_{\mu \nu
}$, the dilaton field $\Phi $ and the gauge field $A_{\mu }$ which
yields the following field equations
\begin{equation}
R_{\mu \nu }=2 \partial _{\mu }\Phi
\partial _{\nu }\Phi+2e^{-2 \alpha \Phi}\left( F_{\mu \eta }F_{\nu }^{\text{
}\eta }-\frac{1}{6}g_{\mu \nu }F_{\lambda \eta }F^{\lambda
\eta }\right) ,  \label{FE1}
\end{equation}
\begin{equation}
\nabla ^{2}\Phi =-\frac{\alpha }{2}e^{-2\alpha
\Phi}F_{\lambda \eta }F^{\lambda \eta },  \label{FE2}
\end{equation}
\begin{equation}
\partial_{\mu}{\left(\sqrt{-g} e^{-2\alpha \Phi}F^{\mu \nu }\right)}=0. \label{FE3}
\end{equation}
We would like to find rotating solutions of the above field
equations. For small rotation, we can solve
Eqs.(\ref{FE1})-(\ref{FE3}) to the first order in the angular
momentum parameter $a$.
Inspection of the slowly rotating black hole 
solutions \cite{ShAll}
shows that the only terms in the metric that change to the first
order of the angular momentum parameter $a$ are $g_{t\phi}$ and $g_{t\psi}$.
Similarly, the dilaton field does not change to $O(a)$ and,
$A_{\phi}$ and $A_{\psi}$ are the only components of the vector potential that
change.
Therefore, for an infinitesimal angular momentum, we assume
the metric, the gauge potential and the dilaton field being of the following forms
\begin{eqnarray}\label{metric1}
ds^2 =-u(r)dt^2+h(r) \left[ w(r)dr^2 + \frac{r^2}{4}
\left\{ k(r)d\Omega_{S^{2}}^2 + \sigma_{3}^2 + 2 a f(r) dt \sigma _3 \right\}
\right],
\end{eqnarray}
\begin{eqnarray}
A &=& \frac{\sqrt{3}\sinh{(\vartheta)}\cosh{(\vartheta)}}{\sqrt{4+3\alpha^2}}
\frac{r_{\infty}^2r_{+}^2(r_{+}^2-r_{\infty}^2)}
{(r_{+}^2(r_{\infty}^2-r^2)\sinh^2{(\vartheta)}+(r_{\infty}^2 -r_{+}^2)r^2)
(r_{+}^2\cosh^2{(\vartheta)}-r_{\infty}^2)} \nonumber
\\
& & \times \left( dt-\frac{a}{2}\sigma_{3} \right),  \label{A1}
\end{eqnarray}
\begin{eqnarray}\label{Phi1}
\Phi(r)=\frac{-3\alpha}{4+3\alpha^2}\ln
{\left(1+\frac{r_{+}^2}{r_{\infty}^2-r_{+}^2}\frac{r_{\infty}^2-r^2}{r^2}\sinh^2(\vartheta)\right)},
\end{eqnarray}
where $\sigma_3 = d\psi + \cos \theta d\phi$, $d\Omega_{S^{2}}^2
=d\theta^{2}+\sin^2\theta d\phi^{2}$ denotes the metric of the
unit two-sphere, $r_+$ and $r_\infty$ are constants. The functions
$u(r)$, $h(r)$, $w(r)$, $k(r)$ and $f(r)$ should be determined. In
the particular case, $a=0$, this metric (\ref{metric1}) reduces to
the static Kaluza-Klein dilaton black hole solutions \cite{SSY}.

Here, we are looking for the asymptotically locally flat solutions in the case
$a\neq0$. Our strategy for obtaining the solution is the
perturbative method 
proposed by Horne and Horowitz \cite{Hor1}. For small $a$, we can
expect to have solutions with $u(r)$, $h(r)$, $w(r)$ and $k(r)$
still functions of $r$ alone. Inserting metric (\ref{metric1}),
the gauge potential (\ref{A1}) and the dilaton field (\ref{Phi1})
into the field equations (\ref{FE1})-(\ref{FE3}), one can show
that the static part of the metric leads to the following
solutions \cite{SSY}
\begin{eqnarray}\label{u}
u(r) = \frac{1-\frac{r_{+}^2}{r^2}}{\left(1-\frac{r_{+}^2}{r_{\infty}^2}\right)h^2(r)} ,
\end{eqnarray}
\begin{eqnarray}\label{h}
h(r) =\left[ 1+\frac{r_{+}^2}{r_{\infty}^2-r_{+}^2}\frac{r_{\infty}^2-r^2}{r^2}\sinh^2(\vartheta)\right]
^{\frac{4}{4+3\alpha^2}} ,
\end{eqnarray}
\begin{eqnarray}\label{w}
w(r) = \frac{(r_{\infty}^2-r_{+}^2)^2r_{\infty}^4}{(r_{\infty}^2-r^2)^4 \left( 1-\frac{r_{+}^2}{r^2} \right)} ,
\end{eqnarray}
\begin{eqnarray}\label{k}
k(r) = \frac{(r_{\infty}^2-r_{+}^2)r_{\infty}^2}{(r_{\infty}^2-r^2)^2} ,
\end{eqnarray}
while the rotating part of the metric admits a solution
\begin{eqnarray}\label{f00}
f(r)&=&\frac{1}{h^3(r)r^4(2-3\alpha^2)(r_{\infty}^2-r_{+}^2)(r_{+}^2-r_{\infty}^2+r_{+}^2\sinh^2{(\vartheta)})r_{+}^2\sinh^2{(\vartheta)}} \nonumber
\\
&& \times \left[((6\alpha^2 -4)r_{\infty}^2r_{+}^2+12r_{\infty}^2r^2-(4+3\alpha^2)(r_{\infty}^4+r^4))r_{+}^4\sinh^4{(\vartheta)}\right. \nonumber
\\
&& \left.-(r_{\infty}^2-r_{+}^2)((6\alpha^2 -4)r_{\infty}^2r_{+}^2+12r_{\infty}^2r^2-(8+6\alpha^2)r^4)r_{+}^2\sinh^2{(\vartheta)}\right. \nonumber
\\
&& \left.-(4+3\alpha^2)(r_{\infty}^2-r_{+}^2)^2r^4\right] .
\end{eqnarray}
The coordinates $(t,~r,~\theta,~ \phi,~ \psi )$ run the ranges of
$-\infty < t < \infty ,~ 0 < r < r_\infty ,~ 0 \leq \theta \leq
\pi ,~ 0 \leq \phi \leq 2 \pi$ and $0 \leq \psi \leq 4 \pi$,
respectively. The spacetime (\ref{metric1}) has the timelike
Killing vector field, $\partial / \partial t$, and the spacelike
Killing vector fields with closed orbits, $\partial / \partial
\phi $ and $\partial / \partial \psi $. To avoid the existence of
naked singularities and closed timelike curves on and outside the
black hole horizon, we choose the parameters such that $0 < r_+ <
r_\infty$.

The black hole horizon is located at $r = r_+$. The induced metric
on the three-dimensional spatial cross section of the black hole
horizon with the time slice is obtained as
\begin{eqnarray}
 ds^2 |_{r=r_+,~t=\text{const.}} = \left[ \cosh (\vartheta) \right]^\frac{8}{4+3\alpha^2}
\frac{r_+ ^2}{4}
\left[
k(r_+) d\Omega_{S^{2}}^2 + \sigma_{3}^2
\right] ,
\end{eqnarray}
which implies the shape of horizon is the squashed $S^3$,
a twisted $S^1$ fiber bundle over an $S^2$ base space with the different sizes.
We see that the function $k(r)$ causes the deformation of the black hole horizon.
We also note that the rotation parameter $a$ has no contribution to the shape of the horizon,
in contrast to the rotating squashed Kaluza-Klein black hole solutions in \cite{DM,TW,NIMT}.

\section{ Asymptotic structure of the solutions}
In the coordinate system $(t,r,\theta,\phi,\psi)$, the metric
(\ref{metric1}) diverges at $r=r_{\infty}$, but we see that this
is an apparent singularity and corresponds to the spatial
infinity. In order to see this, we introduce the new radial
coordinate $\rho$ given by
\begin{equation}\label{Rho1}
\rho=\frac{\sqrt{r_{\infty}^2-r_{+}^2}}{2}\frac{r^2}{r_{\infty}^2-r^2}.
\end{equation}
We also define the parameters
\begin{equation}\label{Rho2}
\rho_{+} = \rho{(r_{+})} = \frac{r_{+}^2}{2\sqrt{r_{\infty}^2-r_{+}^2}},
\end{equation}
\begin{equation}\label{Rho3}
\rho_{0}=\frac{1}{2}\sqrt{r_{\infty}^2-r_{+}^2} .
\end{equation}
In terms of the coordinate $\rho$, the metric, the gauge potential
and the dilaton field can be written as
\begin{eqnarray}\label{metric3}
ds^2 = -U(\rho)dt^2 + V(\rho)d\rho^2 + H(\rho)
\left[ \rho(\rho+\rho_{0}) d\Omega_{S^{2}}^2 + \frac{(\rho_0 + \rho_+)\rho_0 \rho}{\rho +\rho _0} \sigma_{3}^2
+ a F(\rho ) dt \sigma _3\right] ,
\end{eqnarray}
\begin{equation}
A=\frac{\sqrt{3}\sinh{(\vartheta)}\cosh{(\vartheta)}}{\sqrt{4+3\alpha^2}}\frac{\rho_{+}(\rho+\rho_{0})}{(\rho_{0}-\rho_{+}\sinh^2{(\vartheta)})(\rho+\rho_{+}\sinh^2{(\vartheta)})}
\left( dt-\frac{a}{2}\sigma_{3} \right) ,  \label{A2}
\end{equation}
\begin{eqnarray}\label{Phi2}
\Phi(\rho)=\frac{-3\alpha}{4+3\alpha^2} \ln {\left[1+\frac{\rho_{+}}{\rho}\sinh^2{(\vartheta)}\right]} ,
\end{eqnarray}
where the functions $U(\rho)$, $V(\rho)$, $H(\rho)$ and $F(\rho)$ are, respectively, given by
\begin{equation}\label{U}
U(\rho)=\left(1 - \frac{\rho_{+}}{\rho}\right)\frac{1}{H^2(\rho)},
\end{equation}
\begin{equation}\label{V}
V(\rho)=H(\rho)\left(\frac{\rho+\rho_{0}}{\rho-\rho_{+}}\right),
\end{equation}
\begin{equation}\label{H}
H(\rho)=\left[1+\frac{\rho_{+}}{\rho}\sinh^2{(\vartheta)}\right]^{\frac{4}{4+3\alpha^2}},
\end{equation}
\begin{eqnarray}\label{F}
F(\rho)&=&
\frac{1}{2 \left(3 \alpha ^2-2\right)
   \rho  \rho _+ \left(\sinh
   ^2(\vartheta ) \rho _+-\rho
   _0\right) \left(\rho +\rho
   _0\right) H^3 (\rho)}  \nonumber
\\
&& \times
\left[\sinh ^2(\vartheta )
   \left(\left(4-6 \alpha ^2\right)
   \rho +\left(8-3 \alpha ^2\right)
   \rho _0\right) \rho _+^3 \right. \nonumber
\\
&&
+\left(2
   \left(3 \alpha ^2-2\right) \rho ^2
   \sinh ^2(\vartheta )+\left(6
   \alpha ^2+\left(3 \alpha
   ^2+4\right) \sinh ^2(\vartheta
   )-4\right) \rho _0^2 \right. \nonumber
\\
&& \left.
+\rho  \left(9
   \alpha ^2+\left(3 \alpha
   ^2-2\right) \cosh (2 \vartheta
   )+6\right) \rho _0\right) \rho_+^2 \nonumber
\\
&&
+\rho  \rho _0 \left(\rho
   \left(-6 \alpha ^2+\left(3 \alpha
   ^2+4\right)
   \text{csch}^2(\vartheta
   )+4\right)+12 \rho _0\right) \rho_+ \nonumber
\\
&& \left.
+\left(3 \alpha ^2+4\right) \rho
   ^2 \text{csch}^2(\vartheta ) \rho
   _0^2
\right]  .
\end{eqnarray}
The new radial coordinate $\rho$ runs from $0$ to $\infty$. In the
limit $\rho \rightarrow \infty$ (i.e.  $ r\rightarrow
r_{\infty}$), the metric (\ref{metric3}) reduces to
\begin{eqnarray}\label{metric35}
ds^2=-dt^2+d\rho^2+\rho^2d\Omega_{S^{2}}^2
+ \rho_{0} \left( \rho_{0} + \rho_{+} \right) \sigma^2_{3}
+ a F_{\infty} dt\sigma_{3} ,
\end{eqnarray}
where
\begin{equation}
F_{\infty} = F(\infty) =
\frac{\left(3 \alpha ^2+4\right) \rho
   _0 \left(\rho _++\rho _0\right)
   \text{csch}^2(\vartheta )
+ \left(3 \alpha ^2-2\right) \rho _+ \left[ \rho _+ \cosh (2
   \vartheta ) - \rho _+ -2 \rho _0 \right]}
{2 \left(3 \alpha ^2-2\right) \rho _+
   \left(\sinh ^2(\vartheta ) \rho
   _+-\rho _0\right)}  .  \label{FFFF}
\end{equation}
Next, in order to transform the asymptotic frame into the rest
frame, we define the coordinate $\psi_{\ast}$ given by
\begin{equation}\label{Psi}
{\psi_{\ast}}=\psi + \frac{a F_{\infty}}{2 \rho_{0} \left( \rho_{0} + \rho_{+} \right)} t .
\end{equation}
Then, the metric takes the following asymptotic form
\begin{eqnarray}\label{metric4}
ds^2=-dt^2+d\rho^2+\rho^2d\Omega_{S^{2}}^2
+ L^2 {\sigma_{\ast}}^2_{3},
\end{eqnarray}
where
\begin{equation}\label{tidlesigma}
{\sigma_{\ast}}_{3} = d{\psi_{\ast}} + \cos{\theta} d\phi ,
\end{equation}
and the size of the extra dimension $L$ is given by
$L^2 = \rho_{0} \left( \rho_{0} + \rho_{+} \right) = r_\infty ^2 / 4$.
 We see that the spacetime is asymptotically locally flat,
i.e., the asymptotic form of the metric is a twisted $S^1$ bundle
over the four-dimensional Minkowski spacetime.

\section{Various Limits}
\subsection{$\alpha \to 0$}
One may note that in the absence of a non-trivial dilaton, $\alpha
= 0$, solution (\ref{metric1}) reduces to the slowly rotating
charged squashed Kaluza-Klein black hole solutions of the
five-dimensional Einstein-Maxwell theory with two equal angular
momenta. To see this, we introduce a coordinate $\tilde{r}$ such
that
\begin{equation}\label{rr}
\tilde{r}^2=r^2+\frac{r_{+}^2}{r_{\infty}^2-r_{+}^2}(r_{\infty}^2-r^2)\sinh^2(\vartheta),
\end{equation}
and the new parameters
\begin{equation}\label{rr+}
\tilde{r}_{+}=r_{+} \cosh(\vartheta),
\end{equation}
\begin{equation}\label{rr-}
\tilde{r}_{-}=\frac{r_{+}r_{\infty}}{\sqrt{r_{\infty}^2-r_{+}^2}}\sinh(\vartheta),
\end{equation}
\begin{equation}\label{rrinfty}
\tilde{r}_{\infty}=r_{\infty}.
\end{equation}
Therefore, the metric and the gauge potential reduce to
\begin{eqnarray}\label{metric0}
ds^2 = -\frac{\breve{w}(\tilde r)}{\breve{w}(\tilde r_{\infty})}dt^2
+ \breve{k}(\tilde r)^2\frac{d\tilde r^2}{\breve{w}(\tilde r)}
+ \frac{\tilde r^2}{4}
\left[
\breve{k}(\tilde r)d\Omega_{S^{2}}^2 + \sigma_{3}^2 + 2a \breve{f}(\tilde r) dt\sigma _3
\right],
\end{eqnarray}
\begin{equation}
\breve{A}=\frac{\sqrt{3} \tilde r_{+} \tilde r_{-}}{2 \tilde r^2\sqrt{\breve{w}( \tilde r_{\infty})}}
\left( dt-\frac{a}{2}\sigma_{3} \right),  \label{A0}
\end{equation}
where the functions $\breve{w}(\tilde r),~\breve{k}(\tilde r)$ and
$\breve{f}(\tilde r)$ are now given by
\begin{eqnarray}
\breve{w}(\tilde r) =\frac{(\tilde r^2- \tilde r_{+}^2)( \tilde r^2- \tilde r_{-}^2)}{ \tilde r^4}
,  \label{w0}
\end{eqnarray}
\begin{eqnarray}
\breve{k}(\tilde r) =\frac{(\tilde r_{\infty}^2-\tilde r_{+}^2)(\tilde r_{\infty}^2-\tilde r_{-}^2)}
{(\tilde r_{\infty}^2 - \tilde r^2)^2}
,  \label{k0}
\end{eqnarray}
\begin{eqnarray}
\breve{f}(\tilde r) =
\frac{2\tilde r_{\infty}^4(\tilde r_{+}^2 \tilde r_{-}^2-(\tilde r_{+}^2+\tilde r_{-}^2)\tilde r^2)}
{(\tilde r_{\infty}^2-\tilde r_{+}^2)(\tilde r_{\infty}^2-\tilde r_{-}^2)\tilde r^6} ,  \label{f0}
\end{eqnarray}
To avoid the existence of the naked singularities and closed
timelike curves on and outside the black hole horizon, we choose
the parameters such that $0 < \tilde r_- < \tilde r_+ < \tilde
r_\infty$.
For $ \tilde r_{\infty} \rightarrow \infty$,
the solution (\ref{metric0}) is just
the five-dimensional slowly rotating Kerr-Newman black hole
with two equal angular momenta \cite{Aliev2}.
When the rotation parameter vanishes, $a \to 0$,
the solution (\ref{metric0}) coincides with
the five-dimensional charged static Kaluza-Klein black hole
with squashed horizons \cite{IM}.
For $\tilde r_{-} \rightarrow 0$, the solution (\ref{metric0}) reduces to
the vacuum rotating squashed Kaluza-Klein black hole in the limit of slow rotation
\cite{DM,TW}.
\subsection{$r_\infty \to \infty$}
For $ r_{\infty}\rightarrow \infty$, solution (\ref{metric1})
reduces to the five-dimensional slowly rotating charged dilaton
black hole solution with two equal angular momenta
\cite{ShAll}. To show this, we define the new parameters
\begin{equation}\label{alpha}
\tilde{\alpha}=\sqrt{\frac{3}{2}}\alpha,
\end{equation}
\begin{equation}\label{a}
\tilde{a}=\frac{a}{2},
\end{equation}
then, the metric reduces to  
\begin{equation}\label{AFmet}
 ds^2 = - \tilde U(\tilde r) dt^2 + \frac{d\tilde r^2}{\tilde W (\tilde r)}
+ \tilde r^2 \tilde R^2 (\tilde r) d\Omega_{S^3}^2
+ \tilde a \tilde F(\tilde r) dt \sigma_3 ,
\end{equation}
where $d\Omega_{S^3}^2 = \frac{1}{4} \left( d\Omega_{S^2}^2 +
\sigma_3 ^2 \right)$ denotes the metric of the unit three-sphere,
and the functions $\tilde U(\tilde r),~\tilde W(\tilde r),~\tilde
R(\tilde r)$ and $\tilde F(\tilde r)$ are now given by
\begin{equation}\label{AFU}
 \tilde U(\tilde r) = \left( 1-\frac{\tilde r_+^2}{\tilde r^2} \right)
 \left( 1-\frac{\tilde r_-^2}{\tilde r^2} \right)^\frac{2-\tilde \alpha ^2}{2+\tilde \alpha ^2} ,
\end{equation}
\begin{equation}\label{AFW}
 \tilde W(\tilde r) = \left( 1-\frac{\tilde r_+^2}{\tilde r^2} \right)
 \left( 1-\frac{\tilde r_-^2}{\tilde r^2} \right)^\frac{2}{2+\tilde \alpha ^2} ,
\end{equation}
\begin{equation}\label{AFR}
 \tilde R(\tilde r) = \left( 1-\frac{\tilde r_-^2}{\tilde r^2} \right)^\frac{\tilde \alpha ^2}{2(2+\tilde \alpha ^2)} ,
\end{equation}
\begin{equation}\label{AFF}
 \tilde F(\tilde r) =
\left( 1-\frac{\tilde r_-^2}{\tilde r^2} \right)^\frac{2-\tilde
\alpha ^2}{2+\tilde \alpha ^2} \left[ \frac{2+\tilde \alpha
^2}{1-\tilde \alpha ^2} \frac{\tilde r^2}{\tilde r_- ^2}
\left\{1-\left( 1-\frac{\tilde r_-^2}{\tilde r^2}
\right)^\frac{2(\tilde\alpha ^2-1)}{2+\tilde \alpha ^2}\right\}+2
\left( 1-\frac{\tilde r_+^2}{\tilde r^2}\right) \right].
\end{equation}
The spacetime (\ref{AFmet}) asymptotes to the five-dimensional
Minkowski spacetime at infinity. In the absence of a dilaton
field, $\tilde \alpha = 0$, the metric (\ref{AFmet}) reduces to
Eq. (\ref{metric0}) with $\tilde r _\infty \to \infty$
\cite{Aliev2}. When $\tilde a \to 0$, the solution (\ref{AFmet})
coincides with the five-dimensional charged static dilaton black
hole \cite{HS}.
\subsection{$\rho _0 \to 0$}
We consider the limit $\rho _0 \to 0$ with $\rho _+$ finite. We
introduce the coordinates $(t',~\psi ')$ and the parameter $a'$
defined as
\begin{equation}\label{coordsrho00}
t' = t - \frac{a}{2} \psi , \quad
\psi ' = \sqrt{\rho _0 (\rho _0 +\rho _+)} \psi , \quad
a' = \frac{a}{\sqrt{\rho _0 (\rho _0 +\rho _+)}} .
\end{equation}
Then, the metric (\ref{metric3}) reduces to
\begin{equation}\label{metrho00}
ds^2 = - H^{-2}(\rho ) \left( 1-\frac{\rho _+}{\rho } \right) dt'^2
+ H(\rho ) \left[ \left( 1-\frac{\rho _+}{\rho } \right)^{-1} d\rho ^2
+ \rho ^2 d\Omega _{S^2}^2 + d\psi '^2
\right] ,
\end{equation}
where the function $H(\rho )$ is given by Eq.(\ref{H}).
This metric (\ref{metrho00}) coincides with that of
charged static dilaton black strings \cite{KKR}.
\section{Physical Quantities }
In this section we would like to calculate the mass, the angular
momentum and the gyromagnetic ratio of these rotating Kaluza-Klein
dilaton black holes. Starting with (\ref{metric3}), after a few
calculations, the Komar mass $M$ associated with the timelike
Killing vector field $\partial / \partial t$ at infinity  and the
Komar angular momentums $J_\phi $ and $J_{\psi}$ associated with
the spacelike Killing vector fields $\partial / \partial \phi $
and $\partial / \partial \psi _*$ at infinity can be obtained as
\begin{equation}\label{mass}
M = \frac{3\pi\rho_{+}\sqrt{(\rho_{0}+\rho_{+})\rho_{0}}(3\alpha^2+4\cosh{(2\vartheta)})}{8+6\alpha^2} ,
\end{equation}
\begin{equation}\label{Jphi}
J_{\phi} = 0 ,
\end{equation}
\begin{eqnarray}\label{Jpsi}
J_{\psi} &=& a \pi\rho_{+}\sqrt{\rho_{0}(\rho_{0}+\rho_{+})} \nonumber \\
&& \times \left[\frac{(5-3\alpha^2)\rho_{+}-9\alpha^2\rho_{+}+
((-8+3\alpha^2)\rho_{+}-(16+3\alpha^2)\rho_{0})\cosh{(2\vartheta)}+3\rho_{+}\cosh{(4\vartheta)}}
{2(4+3\alpha^2)(-\rho_{+}-2\rho_{0}+\rho_{+}\cosh{(2\vartheta)})}\right]. \nonumber \\
\end{eqnarray}
We see that the spacetime (\ref{metric3}) has only one angular
momentum in the direction of the extra dimension.
For $ r_{\infty}\rightarrow \infty$, 
in terms of the parameters $\tilde a,~ \tilde r_\pm$ and $\tilde \alpha$,
the mass (\ref{mass}) and the angular momentum (\ref{Jpsi}) reduce to
\begin{eqnarray}\label{Mrinfty}
M &=& \frac{3 \pi
\left[ (2+\tilde{\alpha}^2)\tilde{r}_{+}^2 + (2-\tilde{\alpha}^2) \tilde{r}_{-}^2 \right]}
{8 (2+\tilde{\alpha}^2)},
\end{eqnarray}
\begin{eqnarray}\label{Jrinfty}
J &=& \frac{\pi \tilde{a}
\left[ 2(2+\tilde{\alpha}^2)\tilde{r}_{+}^2 + (4-\tilde{\alpha}^2) \tilde{r}_{-}^2 \right]}
{8 (2+\tilde{\alpha}^2)},
\end{eqnarray}
which are the mass and the angular momentum of
the five-dimensional charged slowly rotating dilaton black hole
with equal rotation parameters \cite{ShAll}.

Next, we calculate the gyromagnetic ratio $g$ of slowly rotating
charged Kaluza-Klein dilaton black holes. The gyromagnetic ratio
is an important characteristic of charged rotating black holes.
Indeed, one of the remarkable facts about a Kerr-Newman black hole
in four-dimensional asymptotically flat spacetime is that it can
be assigned a gyromagnetic ratio $g = 2$, just as an electron in
the Dirac theory. It should be noted that, unlike four dimensions,
the value of the gyromagnetic ratio is not universal in higher
dimensions \cite{Aliev3}. Besides, scalar fields, such as the
dilaton, modify the value of the gyromagnetic ratio of the black
hole and consequently it does not possess the gyromagnetic ratio
$g = 2$ of the Kerr-Newman black hole \cite{Hor1}. In our
solution, we also expect the modification of the gyromagnetic
ratio by the non-trivial Kaluza-Klein asymptotic structure, which
is related to the parameter $r_\infty$.  The magnetic dipole
moment can be defined as
\begin{equation}\label{mu}
{\mu}=Q a ,
\end{equation}
where $Q$ denotes the electric charge of the black hole.
The gyromagnetic ratio is defined as a constant of proportionality
in the equation for the magnetic dipole moment
\begin{equation}\label{mu}
{\mu}=g\frac{QJ}{2M}.
\end{equation}
Substituting $M$ and $J$ from Eqs. (\ref{mass}) and (\ref{Jpsi}),
the gyromagnetic ratio can be obtained as
\begin{eqnarray}\label{g}
g =
\frac{6(3\alpha^2+4\cosh{(2\vartheta)})(-\rho_{+}-2\rho_{0}+\rho_{+}\cosh{(2\vartheta)})}
{\rho_{+}(5-3\alpha^2)-9\alpha^2\rho_{0}+((-8+3\alpha^2)\rho_{+}-(16+3\alpha^2)\rho_{0})
\cosh{(2\vartheta)}+3\rho_{+}\cosh{(4\vartheta)}} .
\end{eqnarray}
In terms of $\tilde \alpha$, $\tilde r_\pm$ and $\tilde r_\infty$,
the gyromagnetic ratio is given by
\begin{eqnarray}\label{g2}
g =
\frac{6 \left[ ((2+\tilde{\alpha}^2)\tilde{r}_\infty ^2 -4\tilde{r}_- ^2) \tilde{r}_{+}^2
+ (2-\tilde{\alpha}^2)\tilde{r}_{-}^2 \tilde{r}_\infty ^2 \right]}
{2((2+\tilde{\alpha}^2)\tilde{r}_\infty ^2 -3\tilde{r}_- ^2) \tilde{r}_{+}^2
+(4-\tilde{\alpha}^2)\tilde{r}_{-}^2 \tilde{r}_\infty ^2} .
\end{eqnarray}
We see that the dilaton field and the Kaluza-Klein asymptotic
structure modify the value of the gyromagnetic ratio of the
five-dimensional Kerr-Newman black hole in the slow rotation limit
\cite{Aliev2} (see also \cite{Kunz3}) through the coupling
parameter $\tilde \alpha$, which measures the strength of the
dilaton-electromagnetic coupling, and the squashing parameter
$\tilde r_\infty$, which is proportional to the size of compact
extra dimension.

\begin{figure}[tbp]
\epsfxsize=12cm
\centerline{\epsffile{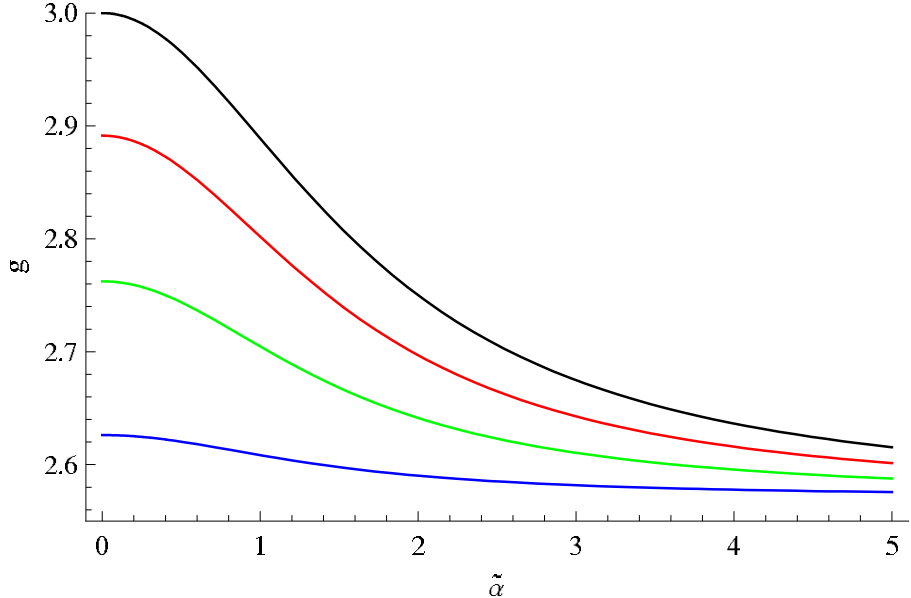}}
\caption{
The behavior of the gyromagnetic ratio $g$ versus $\tilde \alpha$
in various $\tilde r_\infty$ for $\tilde{r}_{+}=2 \tilde{r}_{-}$.
$\tilde r_\infty = 2.1 \tilde{r}_{-}$ (blue line),
$\tilde r_\infty = 2.5 \tilde{r}_{-}$ (green line),
$\tilde r_\infty = 3.5 \tilde{r}_{-}$ (red line), and
$\tilde r_\infty \to \infty$ (black line).
}
\label{figure1}
\end{figure}
In the figure \ref{figure1}, we show the behavior of the
gyromagnetic ratio $g$ versus $\tilde \alpha$ in the range of
parameters $0<\tilde r_- < \tilde r_+ < \tilde r_\infty $.  From
this figure \ref{figure1}, we find out that the gyromagnetic ratio
decreases with increasing $\tilde \alpha$ for  any size of the
compact extra dimension.  In particular, when $r_{\infty}
\rightarrow \infty$,
the gyromagnetic ratio (\ref{g2}) reduces to
\begin{equation}\label{gusuallydilaton}
g = \frac{6\left[ (2+\tilde{\alpha}^2)\tilde{r}_{+}^2+(2-\tilde{\alpha}^2)\tilde{r}_{-}^2 \right]}
{2(2+\tilde{\alpha}^2)\tilde{r}_{+}^2+(4-\tilde{\alpha}^2)\tilde{r}_{-}^2} ,
\end{equation}
which is the gyromagnetic ratio of the five-dimensional slowly
rotating charged dilaton black hole with two equal angular momenta
\cite{ShAll}. Moreover, in the absence of a non-trivial dilaton,
$\tilde \alpha \rightarrow 0$, the gyromagnetic ratio
(\ref{gusuallydilaton}) reduces to
\begin{equation}\label{galiev}
g = 3 ,
\end{equation}
which is the gyromagnetic ratio of the asymptotically flat slowly rotating charged black hole
with two equal angular momenta  \cite{Aliev2}.
\section{Summary and discussion}
To sum up, we have obtained a class of 
slowly rotating charged Kaluza-Klein black hole solutions of
Einstein-Maxwell-dilaton theory with arbitrary dilaton coupling
constant in five dimensions. Our investigations are restricted to
black holes with two equal angular momenta. At infinity, the
metric asymptotically approaches a twisted $S^1$ bundle over the
four-dimensional Minkowski spacetime. Our strategy for obtaining
this solution is the perturbative method proposed by Horne and
Horowitz \cite{Hor1} and solving the equations of motion up to the
linear order of the angular momentum parameter.
We have started from the 
non-rotating charged Kaluza-Klein dilaton black hole solutions in five dimensions
\cite{SSY}.
Then, we have considered the effect of adding a small
amount of rotation parameter $a$ to the solution.
We have discarded any
terms involving $a^2$ or higher powers in $a$.
Inspection of the known
rotating black hole solutions shows that the only terms in the metric which
change to $O(a)$ are $g_{t\phi}$ and $g_{t\psi}$.
Similarly, the dilaton field does not change to $O(a)$.
In the absence of the dilaton field
$(\alpha=0)$, our solution reduces to the slowly rotating charged Kaluza-Klein black
hole solution. 
We have calculated the angular momentum $J$ and the gyromagnetic
ratio $g$ which appear up to the linear order of the angular
momentum parameter $a$. Interestingly enough, we found that the
dilaton field and the Kaluza-Klein asymptotic structure modify the
value of the gyromagnetic ratio $g$ through the coupling parameter
$\alpha$, which measures the strength of the
dilaton-electromagnetic coupling, and the squashing parameter
$r_\infty$, which is proportional to the size of compact extra
dimension. We have also seen that the gyromagnetic ratio crucially
depends on the dilaton coupling constant and decreases with
increasing $\alpha$ for any size of the compact extra dimension.

\acknowledgments{ M. A. would like to thank Jutta Kunz for fruitful discussions
and revision of the paper. K. M. would like to thank Tsuyoshi
Houri, Hideki Ishihara, Masashi Kimura, Takeshi Oota and Yukinori
Yasui for fruitful comments. }



\begin{thebibliography}{999}



\bibitem{Stro} A. Strominger and C. Vafa, Phys. Lett. B {\bf 379}, 99 (1996).

\bibitem{Tan} F. Tangherlini, Nuovo Cimento {\bf 27}, 636 (1963).

\bibitem{Myer} R. C. Myers and M. J. Perry, Ann. Phys. (N.Y.) {\bf 172}, 304
(1986).


\bibitem{Cvetic0} M.~Cvetic and D.~Youm, Phys. Rev. D
{\bf54}, 2612 (1996);\\ D. Youm, Phys. Rep. {\bf316}, 1 (1999).

\bibitem{Cvetic1} M.~Cvetic and D.~Youm,
Nucl.\ Phys.\  B {\bf 477}, 449 (1996).

\bibitem{Cvetic2} M.~Cvetic and D.~Youm, Nucl. Phys. B {\bf 476}, 118
(1996);\\Z. W. Chong, M. Cvetic, H. Lu, and C. N. Pope, Phys. Rev.
D {\bf72}, 041901 (2005).

\bibitem{Aliev2} A. N. Aliev, Phys. Rev. D {\bf 74}, 024011 (2006).

\bibitem{Aliev3}
A. N. Aliev, Mod. Phys. Lett. A {\bf21}, 751 (2006);\\
A. N. Aliev, Phys. Rev. D {\bf75}, 084041 (2007);\\
A. N. Aliev,  Class. Quant. Gravit. {\bf24}, 4669 (2007).

\bibitem{kunz1}  J. Kunz, F. Navarro-Lerida, A. K. Petersen,
Phys. Lett. B {\bf 614}, 104 (2005);\\ J. Kunz, F. Navarro-Lerida and J. Viebahn, Phys. Lett. B 639, 362 (2006);\\ F. Navarro-Lerida, arXiv:0706.0591 [hep-th];\\
H. C. Kim,  R. G. Cai,
Phys. Rev. D {\bf77}, 024045 (2008) .


\bibitem{AlievC} A.N. Aliev and D.K. Ciftci, Phys. Rev. D {\bf 79}, 044004 (2009).


\bibitem{CDB1}  G. W. Gibbons and K. Maeda, Nucl. Phys. B {\bf 298}, 741
(1988);\\ T. Koikawa and M. Yoshimura, Phys. Lett. B {\bf 189}, 29
(1987).

\bibitem{HS}  G.T. Horowitz and A. Strominger, Nucl. Phys. B {\bf 360}, 197 (1991).

\bibitem{CDB2}  D. Garfinkle, G. T. Horowitz and A. Strominger, Phys. Rev. D
{\bf 43}, 3140 (1991);\\ R. Gregory and J. A. Harvey, {\em ibid.}
{\bf 47}, 2411 (1993).

\bibitem{Cai}  R. G. Cai and Y. Z. Zhang,  Phys. Rev. D {\bf 54}, 4891
(1996);\\ R. G. Cai, J. Y. Ji and K. S. Soh, {\em ibid.} {\bf 57},
6547 (1998); \\ R. G. Cai and Y. Z. Zhang, {\em ibid.} {\bf 64},
104015 (2001).

\bibitem{Sheykhi1} A. Sheykhi,  Phys. Rev. D  {\bf76}, 124025
(2007);\\ A. Sheykhi, M. H. Dehghani, N. Riazi, Phys. Rev. D {\bf %
75}, 044020 (2007); \newline A. Sheykhi, M. H. Dehghani, N. Riazi
and J. Pakravan Phys. Rev. D {\bf 74}, 084016 (2006);\newline A.
Sheykhi, N. Riazi, Phys. Rev. D {\bf 75}, 024021 (2007); \\ A.
Sheykhi, Phys. Lett. B {\bf 662}, 7 (2008).


\bibitem{Astefanesei:2006sy}
  D.~Astefanesei, K.~Goldstein and S.~Mahapatra,
  Gen.\ Rel.\ Grav.\  {\bf 40}, 2069 (2008).


\bibitem{Charmousis:2009xr}
  C.~Charmousis, B.~Gouteraux and J.~Soda,
  Phys.\ Rev.\  D {\bf 80}, 024028 (2009).


\bibitem{kun}  H. Kunduri and J. Lucietti, Phys. Lett. B {\bf 609}, 143
(2005); \\ S.S. Yazadjiev Phys. Rev. D {\bf 72}, 104014 (2005).


\bibitem{kunz2}  J. Kunz, D. Maison, F. N. Lerida and J. Viebahn,
Phys. Lett. B {\bf639} (2006) 95.

\bibitem{Bri} Y. Brihaye, E. Radu, C. Stelea, Class. Quant. Gravit. {\bf24}, 4839
(2007).

\bibitem{Cas}  R. Casadio, B. Harms, Y. Leblanc and P. H. Cox, Phys. Rev. D
{\bf 55}, 814 (1997).

\bibitem{Hor1} J. H. Horne and G. T. Horowitz, Phys. Rev. D {\bf 46}, 1340 (1992).


\bibitem{Shi}  K. Shiraishi, Phys. Lett. A {\bf 166}, 298 (1992).

\bibitem{Sheykhi2} A. Sheykhi and N. Riazi, Int. J. Theor. Phys. {\bf 45},
(2006) 2453.
\bibitem{Sheykhi3}  A. Sheykhi and N. Riazi, Int. J. Mod. Phys. A, Vol. {\bf22}, No. 26, (2007)
4849.
\bibitem{Sheykhi4} A. Sheykhi,  Phys. Rev. D {\bf77}, 104022
(2008).


\bibitem{SAcosmo}  A. Sheykhi and M. Allahverdizadeh, Phys. Rev. D {\bf 78}, 064073 (2008).

\bibitem{Setal}  A. Sheykhi, M. Allahverdizadeh, Y. Bahrampour, and M. Rahnama,
Phys. Lett. B {\bf 666}, 82 (2008).

\bibitem{ShAll}
A. Sheykhi and M. Allahverdizadeh, Gen. Rel. Grav. (2009), in press (arXiv:0904.1776 [hep-th]).

\bibitem{DM}
P. Dobiasch and D. Maison, Gen. Rel. Grav. {\bf 14}, 231 (1982).

\bibitem{GW}
G.W. Gibbons and D.L. Wiltshire, Ann. Phys. {\bf 167}, 201 (1986).

\bibitem{Rash2}
D. Rasheed, Nucl. Phys. B {\bf 454}, 379 (1995); \\
T. Matos and C. Mora, Class. Quant. Grav. {\bf 14}, 2331 (1997); \\
F. Larsen, Nucl. Phys. B {\bf 575}, 211 (2000).

\bibitem{GGHPR}
J.~P.~Gauntlett, J.~B.~Gutowski, C.~M.~Hull, S.~Pakis, and H.~S.~Reall, Class. Quant. Grav. {\bf 20}, 4587 (2003).


\bibitem{GSYhole}
D. Gaiotto, A. Strominger and X. Yin, JHEP {\bf 02}, 024 (2006).

\bibitem{IM}
H. Ishihara and K. Matsuno, Prog. Theor. Phys. {\bf 116}, 417 (2006).

\bibitem{IKMTmulti}
H. Ishihara, M. Kimura, K. Matsuno, and S. Tomizawa, Class. Quant. Grav. {\bf 23}, 6919 (2006).

\bibitem{TW}
T. Wang, Nucl. Phys. B {\bf 756}, 86 (2006).

\bibitem{SSY}
S.S. Yazadjiev, Phys. Rev. D {\bf 74}, 024022 (2006).

\bibitem{BR}
Y. Brihaye and E. Radu, Phys. Lett. B {\bf 641}, 212 (2006).


\bibitem{NIMT}
T. Nakagawa, H. Ishihara, K. Matsuno, and S. Tomizawa, Phys. Rev. D {\bf 77}, 044040 (2008).

\bibitem{TIMN}
S.~Tomizawa, H.~Ishihara, K.~Matsuno, and T.~Nakagawa, Prog. Theor. Phys. {\bf 121}, 823 (2009).

\bibitem{MINT}
K.~Matsuno, H.~Ishihara, T.~Nakagawa, and S.~Tomizawa,
  Phys.\ Rev.\  D {\bf 78}, 064016 (2008).

\bibitem{TI}
S.~Tomizawa and A.~Ishibashi,
  Class.\ Quant.\ Grav.\  {\bf 25}, 245007 (2008).

\bibitem{SSW}
C.~Stelea, K.~Schleich and D.~Witt,
  Phys.\ Rev.\  D {\bf 78}, 124006 (2008).

\bibitem{TYM}
S.~Tomizawa, Y.~Yasui and Y.~Morisawa,
  Class.\ Quant.\ Grav.\  {\bf 26}, 145006 (2009).

\bibitem{GS}
D.~V.~Gal'tsov and N.~G.~Scherbluk,
  Phys.\ Rev.\  D {\bf 79}, 064020 (2009).

\bibitem{YIKT}
C.-M.~Yoo, H.~Ishihara, M.~Kimura and S.~Tanzawa,
  arXiv:0906.0689 [gr-qc].


\bibitem{Stelea:2009ur}
  C.~Stelea, K.~Schleich and D.~Witt,
  arXiv:0909.3835 [hep-th].



\bibitem{BKW}
I. Bena, P. Kraus and N.P. Warner, Phys. Rev. D {\bf 72}, 084019 (2005).

\bibitem{EEMRring}
H. Elvang, R. Emparan, D. Mateos and H.S. Reall, JHEP {\bf 08}, 042 (2005).

\bibitem{GSYring}
D. Gaiotto, A. Strominger and X. Yin, JHEP {\bf 02}, 023 (2006).

\bibitem{CEFGS}
J. Camps, R. Emparan, P. Figueras, S. Giusto, and A. Saxena, JHEP {\bf 02}, 021 (2009).

\bibitem{BDGRW}
I. Bena, G. Dall'Agata, S. Giusto, C. Ruef, and N.P. Warner, JHEP {\bf 06}, 015 (2009).



\bibitem{KKR}
B. Kleihaus, J. Kunz, and E. Radu, JHEP {\bf 06}, 016 (2006).


\bibitem{Kunz3}
B. Kleihaus, J. Kunz, E. Radu and C. Stelea, JHEP {\bf 09}, 025 (2009).


\end{thebibliography}
\end{document}